\documentclass[article,preprint,showpacs,amsfonts]{revtex4}
\usepackage{indentfirst}
\usepackage{graphicx}
\usepackage{amsmath,amssymb}
\def\ra{\rangle}
\def\la{\langle}

\begin{document}

\title{Schmidt number criterion via general symmetric informationally complete measurements}

\author{Zhen Wang}
\affiliation {School of Mathematics and Big Data, Jining University, Qufu 273155, China}
\author{Bao-Zhi Sun}
\affiliation {School of Mathematical Sciences, Qufu Normal
University, Qufu 273100, China}
\author{Shao-Ming Fei}
\affiliation {School of Mathematical Sciences, Capital Normal
University, Beijing 100048, China}
\affiliation {Max-Planck-Institute for Mathematics in the Sciences, 04103 Leipzig, Germany}
\author{Zhi-Xi Wang}
\affiliation {School of Mathematical Sciences, Capital Normal
University, Beijing 100048, China}

\begin{abstract}
The Schmidt number characterizes the quantum entanglement of a bipartite mixed state and plays a significant role in certifying entanglement of quantum states. We derive a Schmidt number criterion based on the trace norm of the correlation matrix obtained from the general symmetric informationally complete measurements. The criterion gives an effective way to quantify the entanglement dimension of a bipartite state with arbitrary local dimensions. We show that this Schmidt number criterion is more effective and superior than other criteria such as fidelity, CCNR (computable cross-norm or realignment), MUB (mutually unbiased bases) and EAM (equiangular measurements) criteria in certifying the Schmidt numbers by detailed examples.

\end{abstract}


\maketitle

\section{Introduction}

Quantum entanglement \cite{guhne,horo} is a fundamental feature of quantum systems and plays important roles in various quantum information processing tasks such as quantum communication \cite{bennett,bennett2,buhr}, quantum cryptography \cite{scarani,xu,pira} and quantum teleportation \cite{qt,ot}. Much effort has been devoted to identify the separability and estimate the entanglement of quantum states \cite{plenio,yaohm,jinzx}. One particularly important entanglement measure is the Schmidt number \cite{terhal} or entanglement dimensionality, which quantifies the dimension needed to reproduce the correlations in quantum states \cite{friis}. The detection of the Schmidt rank has significant applications in characterizing entanglement in many-body systems. For example, topologically ordered states with higher Schmidt rank can reflect a more complex entanglement structure.

A quantum state is separable if and only if the Schmidt number is one. The Schmidt number can be used not only to detect entanglement but also to consider how many of their degrees of freedom genuinely exhibit entanglement. Therefore, estimating the bound of the Schmidt number is an effective way to testify different types of entanglement. The first Schmidt number criterion is obtained by examining the fidelity between the quantum state and the maximally entangled states \cite{terhal}. Then in \cite{hulpke,johnston} the authors presented a Schmidt number criterion by generalizing the well-known positive partial transpose \cite{peres,horodecki} and the computable cross-norm or realignment criterion (CCNR) \cite{rudolph,chen} criteria. Similarly, the Schmidt number criterion based on the Bloch representation is derived in \cite{klockl}. Recently, criteria based on covariance matrices have been studied \cite{liu,liu2}. In Ref. \cite{morelli} the Schmidt number criterion is derived by using the sum-total of the probabilities from the global product measurements on a quantum state. In addition, MUBs and equiangular measurements (EAMs) can also be used to witness the Schmidt number via fidelity estimation \cite{morelli,zhang}. More general criteria of Schmidt number have also been given via two well-known families of measurements, the symmetric informationally complete (SIC) measurements and the mutually unbiased bases (MuBs) \cite{tavakoli}. In Ref. \cite{tavakoli} the Schmidt number criterion is based on the trace norm of the correlation matrix whose entries are obtained by measuring each subsystem of a quantum state. In fact, the SIC and MUB criteria for detecting the Schmidt rank can be considered as a generalization of the entanglement criterion in Ref. {\cite{shang}}. The efficiency of the approach has been demonstrated for the entanglement detection, comparing with CCNR criterion.

In this paper, we derive a criterion on the Schmidt number based on the trace norm of the correlation matrix whose entries are obtained via the general symmetric informationally complete (GSIC) measurements instead of SIC measurement. It can be considered as a generalization of the SIC and MUB criteria given in Ref.\cite{tavakoli}. Then it is shown that our criterion gives a more effective way in quantifying the Schmidt number than the existing ones via detailed examples. Finally, we conclude the paper.

\section{GSIC Schmidt number criterion}

A bipartite pure state $|\psi\ra_{AB}=\displaystyle\sum_{ij}c_{ij}|ij\ra\in{\cal H}_{AB}={\cal H}_A\otimes {\cal H}_B$ with ${\rm dim}{\cal H}_A=d_1$ and ${\rm dim}{\cal H}_B=d_2$ has a Schmidt decomposition $|\psi\ra_{AB}=\displaystyle\sum_{i=1}^r\lambda_i|\widetilde{i}\widetilde{i}\ra$, where $\lambda_i>0$ and $\displaystyle\sum_{i=1}^r\lambda_i^2=1$, $\{{|\widetilde{i}}\ra_A\}$ and $\{{|\widetilde{i}}\ra_B\}$ are orthonormal bases in ${\cal H}_A$ and ${\cal H}_B$, respectively. The nonzero number $r$ is called the Schmidt rank of $|\psi\ra$, denoted as ${\rm SR}(|\psi\ra)=r$. For a mixed state $\rho_{AB}\in{\cal H}_{AB}$, the Schmidt rank is generalized to the Schmidt number, denoted as ${\rm SN}(\rho_{AB})$ \cite{terhal}. The Schmidt number of a mixed state $\rho_{AB}$ is the minimization taking over all the pure state decompositions of $\rho_{AB}=\sum_i p_i|\psi_i\ra\la\psi_i|$, that is,
$$
{\rm SN}(\rho_{AB})=\displaystyle\min_{\rho_{AB}=\sum_i p_i|\psi_i\ra\la\psi_i|}\max_i {\rm SR}(|\psi_i\ra).
$$


A set of $d^2$ positive semidefinite operators $\{P_\alpha\}_{\alpha=1}^{d^2}$ in Hilbert space  ${\cal H}_d$ with dimension $d$ is said to be a GSIC positive-operator-values measure (GSIC-POVM) if \cite{apple,kalev}
\begin{eqnarray}
&\displaystyle\sum_{\alpha=1}^{d^2}P_\alpha=\mathbb{I},\\[2mm]
&{\rm tr}(P_\alpha^2)=a\label{equal},\\[2mm]
&{\rm tr}(P_\alpha P_\beta)=\frac{1-ad}{d(d^2-1)},\label{nequal}
\end{eqnarray}
where $\alpha,\beta\in\{1,2,\cdots,d^2\},\ \alpha\neq\beta,$ the parameter $a$ satisfies $\frac{1}{d^3}< a\leq\frac{1}{d^2}$, $\mathbb{I}$ is the identity operator in ${\cal H}_d$ and ${\rm tr}(P_\alpha)=\frac{1}{d}$. Note that $a=\frac{1}{d^2}$ if and only if all $P_\alpha$ are rank one, which gives rise to a SIC-POVM \cite{renes}. The measurement operators of GSIC-POVM can be explicitly constructed \cite{kalev,siudzi}. Let $\{G_0=\frac{1}{\sqrt{d}}\mathbb{I};\ G_\alpha, \alpha=1,2, \cdots,d^2-1\}$ be an orthonormal Hermitian operator basis with traceless $G_\alpha$. Then the following operators form a GSIC-POVM,
\begin{eqnarray*}
P_\alpha&=&\frac{1}{d^2}\mathbb{I}+t[G-d(d+1)G_\alpha], \alpha=1,2,\cdots,d^2-1,\\[2mm]
P_{d^2}&=&\frac{1}{d^2}\mathbb{I}+t(d+1)G,
\end{eqnarray*}
where $G=\sum_{\alpha=1}^{d^2}G_\alpha$. The parameter $t$, which should be so chosen such that $P_\alpha\neq0$, belongs to the range
$$
-\frac{1}{d^2}\frac{1}{\lambda_{\rm max}}\leq t\leq \frac{1}{d^2}\frac{1}{|\lambda_{\rm min}|},
$$
where $\lambda_{\rm max}$ and $\lambda_{\rm min}$ are the maximal and minimal eigenvalue among all the eigenvalues of $G-d(d+1)G_\alpha$, $\alpha=1,2,\cdots,d^2-1$ and $(d+1)G$. $t$ and $a$ satisfy the following relation,
$$
a=\frac{1}{d^3}+t^2(d-1)(d+1)^3.
$$

Denote $I(\sigma)$ the index of coincidence for a POVM $\{P_\alpha\}_{\alpha=1}^N$ with respect to a quantum state $\sigma$,
$$
I(\sigma)\equiv\sum_{\alpha=1}^N|{\rm tr}(P_\alpha\sigma)|^2.
$$
We have the following conclusion.

{\bf Lemma 1}\quad For a GSIC-POVM $\{P_\alpha\}_{\alpha=1}^{d^2}$ and any linear operator $\sigma$ in ${\cal H}_d$, we have
\begin{equation}\label{lemma1}
I(\sigma)=\frac{(ad^3-1){\rm tr}(\sigma\sigma^{\dagger})+d(1-ad)|{\rm tr}\sigma|^2}{d(d^2-1)}.
\end{equation}

{\bf Proof.}\quad By using the GSIC-POVM operators $\{P_\alpha\}_{\alpha=1}^{d^2}$, we construct an orthonormal basis,
\begin{equation}\label{falpha}
F_\alpha=\displaystyle\sqrt{\frac{d(d^2-1)}{ad^3-1}}P_\alpha+\frac{1}{d\sqrt{d}}
(1-\sqrt{\frac{d^2-1}{ad^3-1}})\mathbb{I}.
\end{equation}
Therefore, for any linear operator $\sigma$ we have
\begin{eqnarray}
\sigma&=&\displaystyle\sum_{\alpha=1}^{d^2}{\rm tr}(F_\alpha\sigma)F_\alpha\nonumber\\[2mm]
&=&\displaystyle\sum_{\alpha=1}^{d^2}\frac{d(d^2-1)}{ad^3-1}{\rm tr}(P_\alpha\sigma)P_\alpha-\frac{d-ad^2}{ad^3-1}{\rm tr}\sigma\cdot\mathbb{I}.\label{trace}
\end{eqnarray}
Namely,
\begin{equation}\label{sigma}
\sigma+\frac{d-ad^2}{ad^3-1}{\rm tr}\sigma\cdot\mathbb{I}=\displaystyle\sum_{\alpha=1}^{d^2}\frac{d(d^2-1)}{ad^3-1}{\rm tr}(P_\alpha\sigma)P_\alpha.
\end{equation}
Multiplying each side by $\sigma^\dagger+\frac{d-ad^2}{ad^3-1}{\rm tr}\sigma^\dagger\cdot \mathbb{I}$ and taking the trace, from (\ref{equal}) and (\ref{nequal}) we obtain (\ref{lemma1}).
$\hfill\Box$

For an arbitrary state $\rho$, the index of coincidence is upper bounded by $\frac{ad^2+1}{d(d+1)}$ as $\rho$ is hermite and ${\rm tr}(\rho^2)\leq1$.

Let $\{P_\alpha^A\}_{\alpha=1}^{d_1^2}$ and $\{P_\beta^B\}_{\beta=1}^{d_2^2}$ be two GSIC-POVMs on subsystems $A$ and $B$ of $\rho_{AB}$, respectively. According to Born's rule, we obtain the resulting outcome statistics,
$$
{\mathcal P}_{\alpha\beta}^{{\rm GSIC}}={\rm tr}(\rho_{AB}P_\alpha^A\otimes P_\beta^B).
$$
In the following we denote ${\mathcal P}$ the $d_1^2\times d_2^2$ matrix given by the entries ${\mathcal P}_{\alpha\beta}^{{\rm GSIC}}$. Concerning the trace norm of ${\mathcal P}$, we have the following Schmidt number criterion.

{\bf Theorem 1} (GSIC criterion)\quad For any bipartite state $\rho_{AB}\in{\cal H}^A_{d_1}\otimes {\cal H}^B_{d_2}$, if the Schmidt number of $\rho_{AB}$ is at most $r$, it holds that
\begin{equation}
\|{\mathcal P}\|_{\rm tr}\leq\frac{M}{K}+(r-1)\frac{N}{K},
\end{equation}
where
\begin{eqnarray}
K&=&\sqrt{d_1d_2(d_1^2-1)(d_2^2-1)},\nonumber\\[2mm]
M&=&\sqrt{(a_1d_1^2+1)(a_2d_2^2+1)(d_1-1)(d_2-1)},\nonumber\\[2mm]
N&=&\sqrt{(a_1d_1^3-1)(a_2d_2^3-1)}.\nonumber
\end{eqnarray}

{\bf Proof.}\quad Without loss of generality, since the trace norm is convex, we consider pure states with Schmidt rank $r$, instead of quantum states with Schmidt number $r$,
$|\psi\ra=\displaystyle\sum_{s=1}^r\lambda_s|ss\ra$, where $\{\lambda_s\}$ is the set of Schmidt coefficients with $\displaystyle\sum_{s=1}^r\lambda_s^2=1$.
Thus the probabilities are then given by
\begin{eqnarray}
P_{\alpha\beta}^{\rm GSIC}&=&\displaystyle\sum_{s,t=1}^r\lambda_s\lambda_t\la ss|P_\alpha^A\otimes P_\beta^B|tt\ra\nonumber\\[2mm]
&=&\displaystyle\sum_{s=1}^r\lambda_s^2D_s^{\alpha,\beta}+\displaystyle\sum_{s\neq t}\lambda_s\lambda_tO_{s,t}^{\alpha,\beta},\label{pab}
\end{eqnarray}
where $D_s^{\alpha,\beta}=\la ss|P_\alpha^A\otimes P_\beta^B|ss\ra$ and $O_{s,t}^{\alpha,\beta}=\la ss|P_\alpha^A\otimes P_\beta^B|tt\ra.$

Denote $D_s=\sum_{\alpha,\beta}D_s^{\alpha,\beta}|\alpha\ra\la\beta|$, which can be written as
$D_s=|\Phi_s\ra\la \Psi_s|$, where $|\Phi_s\ra=\sum_\alpha\la s|P_\alpha^A|s\ra|\alpha\ra$ and $|\Psi_s\ra=\sum_\beta\la s|P_\beta^B|s\ra|\beta\ra$. The trace norm of $D_s$ is given by
\begin{eqnarray}
\|D_s\|_{\rm tr}&=&\sqrt{\la\Phi_s|\Phi_s\ra}\sqrt{\la\Psi_s|\Psi_s\ra}\nonumber\\[2mm]
&=&\sqrt{I_A(|s\ra\la s|)}\sqrt{I_B(|s\ra\la s|)},
\end{eqnarray}
where we have used the fact that $\la\Phi_s|\Phi_s\ra$ and $\la\Psi_s|\Psi_s\ra$ yield the index of coincidence for $\{P_\alpha^A\}$ and $\{P_\beta^B\}$, respectively.
According to Lemma 1, we obtain
$$
\|D_s\|_{\rm tr}=\sqrt{\frac{a_1d_1^2+1}{d_1(d_1+1)}}\sqrt{\frac{a_2d_2^2+1}{d_2(d_2+1)}}.
$$

Similarly, for the trace norm of the matrix $O_{s,t}=\sum_{\alpha,\beta}O_{s,t}^{\alpha,\beta}|\alpha\ra\la\beta|$, we have
\begin{eqnarray}
\|O_{s,t}\|_{\rm tr}&=&\sqrt{I_A(|s\ra\la t|)}\sqrt{I_B(|s\ra\la t|)}\nonumber\\[2mm]
&=&\sqrt{\frac{a_1d_1^3-1}{d_1(d_1^2-1)}}\sqrt{\frac{a_2d_2^3-1}{d_2(d_2^2-1)}},
\end{eqnarray}
where the last equality is obtained by using Lemma 1 with the linear operator $\sigma=|s\ra\la t|$.

From Eq. (\ref{pab}), we obtain finally
\begin{eqnarray}
\|{\mathcal P}\|_{\rm tr}&\leq&\displaystyle\sum_{s=1}^r\lambda_s^2\|D_s\|_{\rm tr}+\sum_{s\neq t}\lambda_s\lambda_t\|O_{s,t}\|\nonumber\\[2mm]
&=&\displaystyle\sum_{s=1}^r\lambda_s^2\sqrt{\frac{a_1d_1^2+1}{d_1(d_1+1)}}\sqrt{\frac{a_2d_2^2+1}{d_2(d_2+1)}}+\sum_{s\neq t}\lambda_s\lambda_t\sqrt{\frac{a_1d_1^3-1}{d_1(d_1^2-1)}}\sqrt{\frac{a_2d_2^3-1}{d_2(d_2^2-1)}}\nonumber\\[2mm]
&=&\frac{1}{K}\Big[\displaystyle\big(\sum_{s=1}^r\lambda_s^2\big)\big(M-N\big)+\big(\sum_{s=1}^r\lambda_s\big)^2N\Big]\nonumber\\[2mm]
&\leq&\frac{M+(r-1)N}{K}.
\end{eqnarray}
In the last inequality, we have used the fact that $\displaystyle\big(\sum_{s=1}^r\lambda_s\big)^2\leq r$ \cite{terhal}.
$\hfill\Box$

{\bf Remark 1}\quad When $a_1=\frac{1}{d_1^2},\ a_2=\frac{1}{d_2^2}$, the GSIC-POVMs reduce to SIC-POVMs. Then the GSIC criterion reduces to the Result 1 (SIC criteiron) given in Ref. \cite{tavakoli}.

{\bf Remark 2}\quad A quantum state $\rho_{AB}$ is separable when its Schmidt number $r=1$. Theorem 1 covers the entanglement criterion introduced in Ref. \cite{lai}.

A set of $(d^2-1)\times(d^2-1)$ real orthogonal matrices of the group $O(d^2-1)$ defines a one parameter GSIC-POVM. Different sets of orthogonal matrices give rise to different GSIC POVMs \cite{kalev}. It can be shown that the correlation matrix $\|{\mathcal P}\|_{\rm tr}$ is independent of the choices of GSIC-POVMs. In fact, from a given orthonormal basis $\{P_\alpha\}$, other orthonormal bases $\{\tilde{P}_\alpha\}$ of ${\cal H}_d$ can be obtained from $\{P_\alpha\}$ via
$\tilde{P}_\alpha=\sum_\beta O_{\alpha\beta}P_\beta$, where $O$ is a real orthogonal matrix given by entries $O_{\alpha\beta}$. Correspondingly we have
$$
[\tilde{\mathcal{P}}]_{\alpha\beta}={\rm tr}(\rho_{AB}\tilde{P}_\alpha^A\otimes \tilde{P}_\beta^B)=\displaystyle\sum_{\lambda,\mu=1}^{d^2}O_{\alpha\lambda}^A
O_{\beta\mu}^B[\mathcal{P}]_{\lambda\mu}=O^A \mathcal{P}(O^B)^T.
$$
Therefore, ${\rm tr}(P_\alpha^{A/B}P_\beta^{A/B})={\rm tr}(\tilde{P}_\alpha^{A/B}\tilde{P}_\beta^{A/B})$ and thus $\|\tilde{\mathcal{P}}\|_{\rm tr}=\|{\mathcal P}\|_{\rm tr}$. Hence, $\|{\mathcal P}\|_{\rm tr}$ is independent of the choices of GSIC-POVMs.

In particular, for any bipartite state $\rho_{AB}\in{\cal H}^A_d\otimes {\cal H}^B_d$, if the Schmidt number of $\rho_{AB}$ is at most $r$, it holds that
\begin{equation}\label{coro}
\|{\mathcal P}\|_{\rm tr}\leq\frac{ad^2+1}{d(d+1)}+(r-1)\frac{ad^3-1}{d(d^2-1)}.
\end{equation}

\section{Examples}

Let us consider some examples to illustrate the effectiveness and superiority of the GSIC criterion in the Schmidt number detection. Let $\{P_\alpha\}_{\alpha=1}^{d^2}$ be a set of GSIC-POVM in ${\cal H}^d$ with the parameter $a$. Let $P_\alpha^*$ denote the complex conjugation of $P_\alpha$. Thus $P_\alpha^*$ is another set of GSIC-POVM with the same parameter $a$.

{\bf Example 1}\quad Consider the family of the $3\times3$ bound entangled states
\begin{equation}\label{rhox}
\rho^x=\frac{1}{8x+1} \left(
\begin{array}{ccccccccc}
x&0&0&0&x&0&0&0&x\\
0&x&0&0&0&0&0&0&0\\
0&0&x&0&0&0&0&0&0\\
0&0&0&x&0&0&0&0&0\\
x&0&0&0&x&0&0&0&x\\
0&0&0&0&0&x&0&0&0\\
0&0&0&0&0&0&\frac{1+x}{2}&0&\frac{\sqrt{1-x^2}}{2}\\
0&0&0&0&0&0&0&x&0\\
x&0&0&0&x&0&\frac{\sqrt{1-x^2}}{2}&0&\frac{1+x}{2}
\end{array}
\right)
\end{equation}
in Ref. \cite{horodecki1}. These bound entangled states for $0<x<1$ can be detected by the CCNR criterion, i.e., the Schmidt number of $\rho_x$ is greater than 1. Now we consider the mixtures of $\rho_x$ with the white noise:
$$
\rho(x,q)=q\rho_x+\frac{(1-q)}{9}\mathbb{I},\ 0\leq q\leq 1.
$$

The following nine matrices
\begin{eqnarray}
P_\alpha&=&\frac{1}{9}\mathbb{I}+t(G_9-12G_\alpha),\  \alpha=1,\ 2,\ \cdots,\ 8,\nonumber\\[2mm]
P_9&=&\frac{1}{9}\mathbb{I}+4tG_9\nonumber
\end{eqnarray}
constitute a GSIC-POVM, where

$$
G_1=\left(
\begin{array}{ccc}
\frac{1}{\sqrt{2}}&0&0\\
0&-\frac{1}{\sqrt{2}}&0\\
0&0&0
\end{array}
\right),
\
G_2=\left(
\begin{array}{ccc}
0&\frac{1}{\sqrt{2}}&0\\
\frac{1}{\sqrt{2}}&0&0\\
0&0&0
\end{array}
\right),
\
G_3=\left(
\begin{array}{ccc}
0&0&\frac{1}{\sqrt{2}}\\
0&0&0\\
\frac{1}{\sqrt{2}}&0&0
\end{array}
\right),
$$

$$
G_4=\left(
\begin{array}{ccc}
0&-\frac{i}{\sqrt{2}}&0\\
\frac{i}{\sqrt{2}}&0&0\\
0&0&0
\end{array}
\right),
\
G_5=\left(
\begin{array}{ccc}
\frac{1}{\sqrt{6}}&0&0\\
0&\frac{1}{\sqrt{6}}&0\\
0&0&-\sqrt{\frac{2}{3}}
\end{array}
\right),
\
G_6=\left(
\begin{array}{ccc}
0&0&0\\
0&0&\frac{1}{\sqrt{2}}\\
0&\frac{1}{\sqrt{2}}&0
\end{array}
\right),
$$

$$
G_7=\left(
\begin{array}{ccc}
0&0&-\frac{i}{\sqrt{2}}\\
0&0&0\\
\frac{i}{\sqrt{2}}&0&0
\end{array}
\right),
\
G_8=\left(
\begin{array}{ccc}
0&0&0\\
0&0&-\frac{i}{\sqrt{2}}\\
0&\frac{i}{\sqrt{2}}&0
\end{array}
\right),
\
G_9=\left(
\begin{array}{ccc}
\frac{1}{\sqrt{2}}+\frac{1}{\sqrt{6}}&\frac{1-i}{\sqrt{2}}&\frac{1-i}{\sqrt{2}}\\
\frac{1+i}{\sqrt{2}}&-\frac{1}{\sqrt{2}}+\frac{1}{\sqrt{6}}&\frac{1-i}{\sqrt{2}}\\
\frac{1+i}{\sqrt{2}}&\frac{1+i}{\sqrt{2}}&-\sqrt{\frac{2}{3}}
\end{array}
\right).
$$

\begin{figure}[!htp]
\begin{center}
\includegraphics[width=0.5\textwidth]{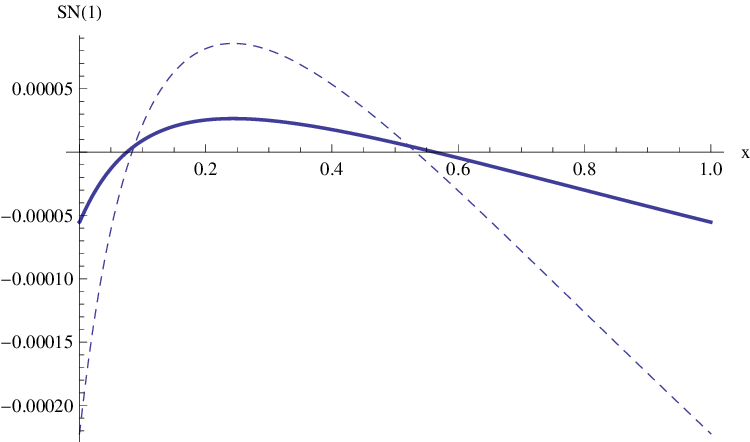}
\caption{The value of $\|{\cal P}\|_\textrm{tr}-\frac{9a+1}{12}$ for the states $\rho(x,q)$ with $q=0.995$. The states above the dashed curve is detected via the SIC-POVMs. The states above the solid curve is detected via the GSIC-POVMs with $t=0.01$.  \label{fig1}}
\end{center}
\end{figure}

We can use the above two GSIC-POVMs $\{P_\alpha\}$ and $\{P_\alpha^*\}$ to recognize the Schmidt number.
In order to compare the GSIC criterion and the SIC criterion, we plot the value of $\|{\cal P}\|_\textrm{tr}-\frac{9a+1}{12}$ for the states $\rho(x,q)$ with $q=0.995$ via the SIC-POVMs (the dashed curve) and the GSIC-POVMs with $t=0.01$ (the solid curve) in FIG.\ref{fig1}. It illustrates the range for which the states $\rho(x,q)$ with $q=0.995$ have Schmidt number greater than 1. It is found that the value of $\|{\cal P}\|_\textrm{tr}-\frac{9a+1}{12}$ via GSIC-POVMs is approximately $1.516\times10^{-6}$ when $x=0.55$. However, the value of $\|{\cal P}\|_\textrm{tr}-\frac{9a+1}{12}$ via SIC-POVMs is approximately $-7.61\times10^{-6}$ when $x=0.55$. From FIG.1 and the numerical results, one can find that the parameter range of $\|{\cal P}\|_\textrm{tr}-\frac{9a+1}{12}>0$ via the GSIC-POVMs with $t=0.01$  is slightly lager than the range via the SIC-POVMs. Thus, our criterion is shown to be more efficient in detecting the Schmidt number of $\rho(x,q)$ than the SIC criterion in Ref. \cite{tavakoli}.

\begin{figure}[!htp]
\begin{center}
\includegraphics[width=0.5\textwidth]{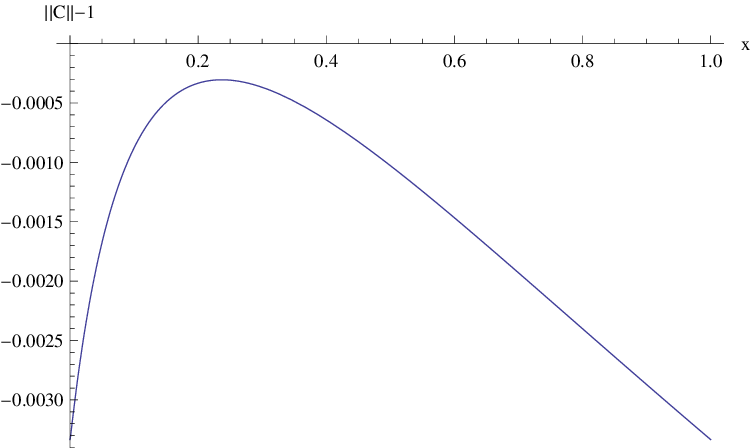}
\caption{The value of $\|C\|_\textrm{tr}-1$ for the states $\rho(x,q)$ with $q=0.995$. \label{fig2}}
\end{center}
\end{figure}

In the following, we compare our criterion with the CCNR criterion. Note that $\{G_0=\frac{1}{\sqrt{3}}\mathbb{I};\ G_\alpha, \alpha=1,2, \cdots,8\}$ constitutes an orthonormal basis of Hermitian operators on ${\cal H}_3$. We plot the value of $\|C\|_\textrm{tr}-1$ for the states $\rho(x,q)$ with $q=0.995$ by use of the correlation matrix
$[C]_{\alpha\beta}={\rm tr}(\rho G_\alpha\otimes G_\beta)$ in FIG.\ref{fig2}. It shows that $\|C\|_\textrm{tr}-1<0$ for all the states $\rho(x,0.995)$. That is, the CCNR criterion fails to detect the states $\rho(x,0.995)$ with the Schmidt number greater than 1. It is obvious that our criterion is more efficient in detecting the Schmidt number of $\rho(x,q)$ than CCNR criterion.

{\bf Example 2}\quad Consider the following isotropic state \cite{horodecki2}
$$
\rho_v=v|\psi^+_d\ra\la\psi^+_d|+\frac{1-v}{d^2}\mathbb{I},
$$
where $|\psi_d^+\ra=\frac{1}{\sqrt{d}}\sum_{i=1}^d|ii\ra$ is the maximally entangled state and $v\in[0,1]$. In \cite{terhal} the authors showed that the Schmidt number of the isotropic state is at least $r+1$ if and only if the visibility $v$ is greater than the critical value $v_{\rm opt}=\frac{rd-1}{d^2-1}$. Below we show that GSIC criterion exactly recovers the optimal visibility thresholds.

Taking into account that $\mathbb{I}\otimes Q|\psi^+_d\ra=Q^{\rm T}\otimes\mathbb{I}|\psi^+_d\ra$ for any linear operator $Q$, we have
\begin{eqnarray}
{\mathcal P}_{\alpha\beta}^{{\rm GSIC}}&=&{\rm tr}(\rho_v P_\alpha\otimes P_\beta^*)\nonumber\\[2mm]
&=&{\rm tr}\big[\big(v|\psi_d^+\ra\la\psi_d^+|+\frac{1-v}{d^2}\mathbb{I}\big)\big(P_\alpha\otimes P_\beta^*\big)\big]\nonumber\\[2mm]
&=&v{\rm tr}(P_\alpha\otimes P_\beta^*|\psi_d^+\ra\la\psi_d^+|)+\frac{1-v}{d^2}{\rm tr}(P_\alpha\otimes P_\beta^*)\nonumber\\[2mm]
&=&v{\rm tr}(P_\alpha P_\beta^\dagger\otimes \mathbb{I}|\psi_d^+\ra\la\psi_d^+|)+\frac{1-v}{d^2}{\rm tr}(P_\alpha){\rm tr}(P_\beta^*)\nonumber\\[2mm]
&=&\frac{v}{d}{\rm tr}(P_\alpha P_\beta)+\frac{1-v}{d^4},\nonumber
\end{eqnarray}
where in the last equality we used the fact that ${\rm tr}(P_\alpha)=\frac{1}{d}$.
As GSIC-POVM satisfies the conditions (\ref{equal}) and (\ref{nequal}), the correlation matrix ${\mathcal P}$ for the state $\rho_v$ can be written as
$$
{\mathcal P}=\frac{v(ad^3-1)}{d^2(d^2-1)}\mathbb{I}+\big[\frac{1}{d^4}+\frac{v(ad^3-1)}{d^4(d^2-1)}\big]J,
$$
where $J$ is the matrix whose entries are all one.

Note that $\lambda$ is an eigenvalue of a matrix $A$ if and only if $\lambda+\mu$ is the eigenvalue of the matrix $A+\mu\mathbb{I}$ \cite{matrix}. The eigenvalues of the correlation matrix ${\mathcal P}$ are $\frac{1}{d^2}$ and $\frac{v(ad^3-1)}{d^2(d^2-1)}$ (with multiplicity $d^2-1$). Then $\|{\mathcal P}\|_{\rm tr}=\frac{1}{d^2}+\frac{v(ad^3-1)}{d^2}$.
For the state $\rho_v$, the fidelity witness given in \cite{terhal} says that,
the isotropic state to have Schmidt number $r+1$ the visibility $v$ must be greater than the critical value $v_{\rm opt}=\frac{rd-1}{d^2-1}$.

In order to compare our criterion with the fidelity witness given in \cite{terhal},
we write $\|{\mathcal P}\|_{\rm tr}$ as
\begin{equation}\label{trace}
\|{\mathcal P}\|_{\rm tr}=\frac{ad^2+1}{d(d+1)}+\frac{ad^3-1}{d^2}(v-\frac{1}{d+1}).
\end{equation}
For any value of $v>v_{\rm opt}$ it holds that
\begin{eqnarray}
\|{\mathcal P}\|_{\rm tr}&>&\frac{ad^2+1}{d(d+1)}+\frac{ad^3-1}{d^2}\big(\frac{rd-1}{d^2-1}-\frac{1}{d+1}\big),\nonumber\\[2mm]
&=&\frac{ad^2+1}{d(d+1)}+(r-1)\frac{ad^3-1}{d(d^2-1)}.\nonumber
\end{eqnarray}
Therefore, our criterion is strictly stronger than the fidelity witness.

In addition, according to (\ref{trace}), the critical visibility for detecting the Schmidt number at least $r+1$ becomes
\begin{equation}
v_{\rm GSIC}=\frac{rd-1}{d^2-1}
\end{equation}
in our criterion. This visibility exactly equals to $v_{\rm opt}$. In Ref. \cite{morelli}, the authors use the sum-total of the probabilities based on MUBs and equiangular measurements (EAMs) to detect the Schmidt number of the quantum state $\rho_v$. It is found that the critical visibility for detecting Schmidt number of $\rho_v$ at least $r+1$ via MUBs and EAMs are $\frac{d-r+m(r-1)}{m(d-1)}$ and $\frac{d-r}{n-1}+\frac{r-1}{d-1}$, respectively. It is clear that our criterion is more superior than the MUB and the EAM criteria in Ref. \cite{morelli}.

{\bf Example 3}\quad Consider the Werner states \cite{werner},
\begin{equation}\label{werner}
\rho_f=\frac{1}{d(d^2-1)}\big[(d-f)\mathbb{I}+(df-1)V\big],
\end{equation}
where $-1\leq f\leq1,\ V=\sum_{i,j=1}^d|ij\ra\la ji|$. $\rho_f$ is entangled if and only if $-1\leq f<0$.

By using that $(I\otimes T)V=d|\psi_d^+\ra\la\psi_d^+|$, where $I,\ T$ are the identity and transposed operator respectively, we have
\begin{eqnarray}
{\mathcal P}_{\alpha\beta}^{{\rm GSIC}}&=&{\rm tr}(\rho_f P_\alpha\otimes P_\beta)\nonumber\\[2mm]
&=&{\rm tr}\big((I\otimes T)(\rho_f P_\alpha\otimes P_\beta)\big)\nonumber\\[2mm]
&=&\frac{1}{d(d^2-1)}{\rm tr}\big[\big((d-f)\mathbb{I}+d(df-1)|\psi_d^+\ra\la\psi_d^+|\big)\big(P_\alpha\otimes P_\beta^*\big)\big]\nonumber\\[2mm]
&=&\frac{1}{d(d^2-1)}\big[(d-f){\rm tr}(P_\alpha\otimes P_\beta^*)+d(df-1){\rm tr}(P_\alpha\otimes P_\beta^*|\psi_d^+\ra\la\psi_d^+|)\big]\nonumber\\[2mm]
&=&\frac{1}{d(d^2-1)}\big[(d-f){\rm tr}(P_\alpha){\rm tr}(P_\beta^*)+d(df-1){\rm tr}(P_\alpha P_\beta^\dagger\otimes \mathbb{I}|\psi_d^+\ra\la\psi_d^+|)\big]\nonumber\\[2mm]
&=&\frac{1}{d(d^2-1)}\big[\frac{d-f}{d^2}+(df-1){\rm tr}(P_\alpha P_\beta)\big].\nonumber
\end{eqnarray}
From the conditions (\ref{equal}) and (\ref{nequal}), the correlation matrix ${\mathcal P}$ for $\rho_f$ can be written as
$$
{\mathcal P}=\frac{1}{d(d^2-1)}\big[\frac{(df-1)(ad^3-1)}{d(d^2-1)}\mathbb{I}+\big(\frac{d-f}{d^2}+\frac{(df-1)(1-ad)}{d(d^2-1)}\big)J\big],
$$
where $J$ is the matrix whose entries are all one. It is verified that the eigenvalues of ${\mathcal P}$ are $\frac{1}{d^2}$ and $\frac{(df-1)(ad^3-1)}{d^2(d^2-1)^2}$ (with multiplicity $d^2-1$).

As ${\mathcal P}$ is symmetric, the singular values of ${\mathcal P}$ equal to the absolute values of its eigenvalues. For $-1\leq f<0$, we have $\|{\mathcal P}\|_{\rm tr}=\frac{1}{d^2}+\frac{(1-df)(ad^3-1)}{d^2(d^2-1)}$. In order to detect the Schmidt number of $\rho_f$, we rewrite $\|{\mathcal P}\|_{\rm tr}$ as
\begin{equation}\label{trace2}
\|{\mathcal P}\|_{\rm tr}=\frac{ad^2+1}{d(d+1)}+(\frac{2}{d}-f-1)\frac{(ad^3-1)}{d(d^2-1)}.
\end{equation}
By using our criterion, the Schmidt number of $\rho_f$ is at least $r+1$ when $f<\frac{2}{d}-r$. Therefore, if $-1\leq f<\frac{2}{d}-1$ we have ${\rm SN}(\rho_f)\geq2$.
That is, our criterion can recognize the entanglement for $-1\leq f<\frac{2}{d}-1$. It coincides with the result in Ref. \cite{lai} for $d=3$.

In the below, it is proved that the entangled states $\rho_f$ have Schmidt number 2. Interestingly, we find a convex decomposition of $\rho_f$:
\begin{eqnarray}
\rho_f&=&\frac{1}{d(d^2-1)}\big[(d-f)\mathbb{I}+(df-1)V\big]\nonumber\\[2mm]
&=&\frac{1}{d(d^2-1)}\big[(d-f)\mathbb{I}+(df-1)(\sum_{i=1}^d|ii\ra\la ii|+\sum_{i<j}|ij\ra\la ji|+|ji\ra\la ij|)\big]\nonumber\\[2mm]
&=&\frac{1}{d(d^2-1)}\big[(d-f)\mathbb{I}+(df-1)\sum_{i=1}^d|ii\ra\la ii|\nonumber\\[2mm]
&+&(1-df)\sum_{i<j}\big(2|\psi_{ij}\ra\la\psi_{ij}|-|ij\ra\la ij|-|ji\ra\la ji|\big)\big]\nonumber\\[2mm]
&=&\frac{1}{d(d^2-1)}\big[(d-1)(1+f)\mathbb{I}+2(1-df)\sum_{i<j}|\psi_{ij}\ra\la\psi_{ij}|\big],
\end{eqnarray}
where $|\psi_{ij}\ra=\frac{1}{\sqrt{2}}(|ij\ra-|ji\ra)$ are pure states with the Schmidt rank 2. The largest Schmidt rank of all the pure states in the convex decomposition is 2. Hence, the entangled Werner states have Schmidt number 2 in deed.

In addition, it is notice that the Schmidt number criterion in Theorem 1 is not necessary and sufficient for pure states.
In fact, let $$|\phi\ra=\frac{1}{5}|00\ra+\frac{1}{5}|11\ra+\frac{\sqrt{23}}{5}|22\ra.$$
It is found that $\|{\cal P}\|_\textrm{tr}<\frac{9a+1}{12}+\frac{27a-1}{24}$ by use of the GSIC-POVMs with $t=0.01$ in Example 1. However, the pure state $|\phi\ra$ has Schmidt rank 3.

\section{Conclusions}
We have developed a criterion for the detection and quantification of high-dimensional entanglement. It can be used to detect the entanglement of bipartite states in arbitrary dimensions. As GSIC measurements cover SIC measurements, our criterion is the generalization of the SIC criterion for the Schmidt number. We have illustrated that our criterion is more effective and superior than the fidelity criterion, CCNR criterion, MUB criterion and EAM criterion in certifying Schmidt number. We have also proved that the entangled Werner states have Schmidt number 2. Our results may highlight further investigations on the detection of quantum entanglement based on other quantum measurements.

\bigskip
\noindent{\bf Acknowledgments}\ The authors would like to thank the referee's valuable suggestions.
This work is supported by Shandong Provincial Natural Science Foundation under Grants ZR2020MA034; the National Natural Science Foundation of China (NSFC) under Grants 12075159 and 12171044; the specific research fund of the Innovation Platform for Academicians of Hainan Province.

\end{document}